\documentclass[prl,amsmath,amssymb, twocolumn, showpacs]{revtex4}
\usepackage{dcolumn}
\usepackage{bm}
\usepackage{graphicx}
\usepackage{hyperref}
\usepackage{mathptmx}
\usepackage{subfigure}
\usepackage[utf8]{inputenc}
\usepackage{bbold}

\newcommand{\RR}{\mathbb{R}}

\newcommand{\Eperp}{E_\bot}
\newcommand{\Lpar}{L_{\|}}
\newcommand{\vv}{\mathbf{v}}

\makeatletter

\begin{document}
\title{Periodizing quasicrystals: Anomalous diffusion in quasiperiodic systems} 
\author{Atahualpa S.~Kraemer}
\email{ata.kraemer@gmail.com}
\author{David P.~Sanders}
\email{dpsanders@ciencias.unam.mx}

\affiliation{Departamento de F\'isica, Facultad de Ciencias, Universidad Nacional Aut\'onoma de M\'exico,
Ciudad Universitaria, 04510 M\'exico D.F., M\'exico.
}

\date{\today}
\begin{abstract}
We introduce a construction to embed a quasiperiodic lattice of obstacles into a single unit cell of a higher-dimensional space,  with \emph{periodic} boundary conditions.
This construction transparently shows the existence of \emph{channels}  in these systems,
in which particles may travel without  colliding, up to a critical obstacle radius. It provides a simple and efficient
algorithm for numerical  simulation of dynamics in quasiperiodic structures, as well as giving a natural notion of uniform distribution (measure) and  averages. 
 As an application, we simulate diffusion in a two-dimensional quasicrystal, finding three different regimes, in particular
 atypical weak super-diffusion in the presence of channels, and sub-diffusion when obstacles overlap. 
\end{abstract}


\maketitle

Quasicrystals are produced by  cooling  from a melt at a rate
intermediate between that of periodic crystals (slow) and  glasses (fast),
and have a degree of order which is  intermediate between the two, being neither periodic nor random
 \cite{67, 68, 74, 13}.
 The transport properties of quasicrystalline materials are of particular interest, with
diffusion
having been extensively studied experimentally  
\cite{104,105,108},
being of importance
 for their production and technological applications \cite{112}. 
This is  related to 
other transport properties, such as heat conductivity and electronic transport 
\cite{104}; furthermore,
 the measured  thermopower in quasicrystals is  due to electron diffusion \cite{109}.

To understand  such transport properties, it is useful to study simple models, in particular  
 the \emph{Lorentz gas} (LG) \cite{84},
which consists of an array of fixed obstacles
in $\RR^{n}$
with which moving particles undergo elastic collisions; 
The geometry in which the obstacles are arranged in such \emph{billiard models} strongly influences these transport properties.

 LGs with a \emph{periodic} geometry have been extensively studied \cite{79,76,77,78,8}.
Usually,  \emph{normal diffusion}
is expected, for which the mean-squared displacement has asymptotic behavior
$ \langle \Delta x(t)^2 \rangle \sim t$ when $t \to \infty$, where $\Delta x(t) := x(t) - x(0)$ is the displacement of a particle at time $t$, and $\langle \cdot \rangle$ denotes 
an average over uniform initial conditions. However, a key role is played by the presence or absence of \emph{channels}
through which particles may travel unimpeded:
in the presence of  ``principal horizons'', i.e.,  channels of the highest possible dimension $n-1$
\cite{8, 100}, there is instead weak super-diffusion with  a logarithmic correction, 
$\langle \Delta x(t)^2 \rangle \sim 
t \ln t$  \cite{89, 8, 90, 100}.

When the obstacles are rather arranged 
randomly,
normal diffusion occurs if the obstacles do not overlap \cite{81}; when overlaps are allowed,  sub-diffusion, with $\langle \Delta x(t)^2 \rangle \sim t^\alpha$ and an exponent $\alpha < 1$, is expected, due to the
presence of arbitrarily large traps near the percolation threshold \cite{82, 117}.

It is then natural to ask what type   of diffusion is expected for the intermediate case of a \emph{quasiperiodic} LG, as was raised in ref.~\cite{16} for 
the special case of  obstacles arranged in a  Penrose tiling. However,  we are not aware of any analytical or numerical results
on such systems, 
except for a rather non-physical one-dimensional model  \cite{111}. 
Indeed, the numerical study of quasiperiodic and random systems is known to be challenging, due 
to the absence of 
periodic boundary conditions, which in the periodic case allow us to restrict attention to a single unit cell. For non-periodic systems, it is often necessary to 
generate at the large structures at the outset, leading to inefficient algorithms \cite{111}.

In this Letter, we introduce a construction to study quasiperiodic systems by 
embedding them in a \emph{periodic} system of higher dimension, which works by reversing 
the projection method used to construct quasiperiodic lattices \cite{95,97}.
This construction solves the above problem 
 for quasiperiodic systems, allowing us to again reduce the system to a single unit cell with periodic boundary conditions, but now in the higher-dimensional 
 system. 
 
 The construction has several consequences: Firstly, it shows in a simple way the existence of channels in quasiperiodic structures, up to a critical radius of the obstacles. Secondly, it 
gives a natural notion of uniform distribution (measure) in the system, and hence of averages, by reducing the infinite system to a finite (compact) one. Finally, it leads directly to a simple and efficient algorithm for simulating dynamics in quasiperiodic structures. 

By applying this simulation method, we  observe three different diffusive regimes for a two-dimensional quasiperiodic LG, including weak super-diffusion in the presence of channels, and sub-diffusion when the obstacles are overlapping.

\paragraph{Projection method:-}
%

The projection method \cite{95,93,98,94,96} constructs a quasiperiodic lattice in a subspace 
$E$  (including the origin)  of a Euclidean space $\RR^n$ by projecting a subset of the vertices of a hypercubic lattice $L$,
consisting of points with integer coordinates in $\RR^n$, onto $E$; for example, the well-known Penrose tiling can be obtained by projecting a 5-dimensional cubic lattice into two dimensions \cite{93, 95, 97}.
We denote by $m$  the dimension of $E$, with $m < n$, and by
 $\Eperp$  the orthogonal complement of $E$, of dimension $n-m$, such that  $E \oplus \Eperp = \RR^n$. 
 In the following, for simplicity, we refer to  orthogonal projections simply as ``projections''.

\begin{figure}
 \centering
  \subfigure[]{
\includegraphics[scale=0.17]{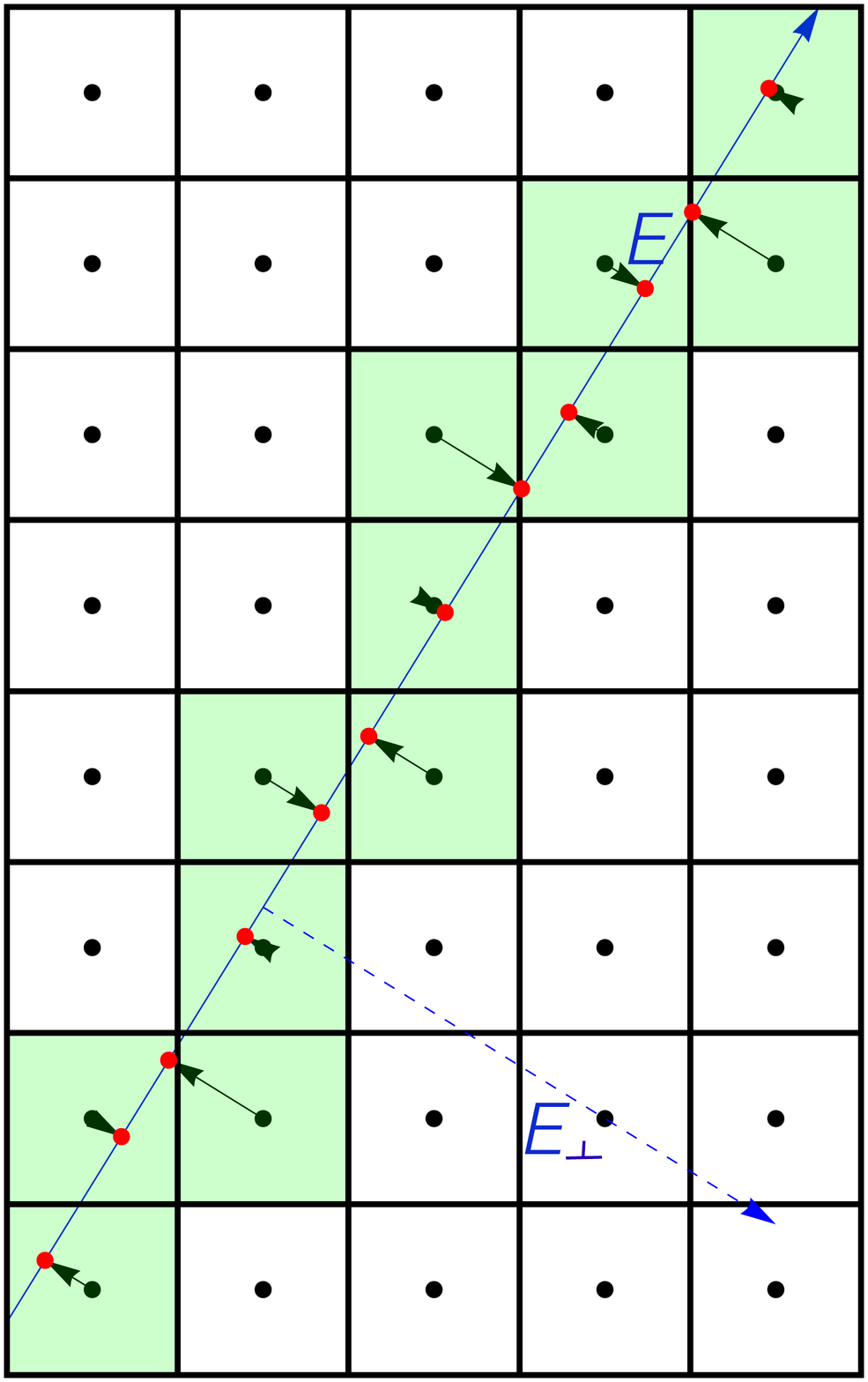}
  }
 \subfigure[]{
 \includegraphics[scale=0.17]{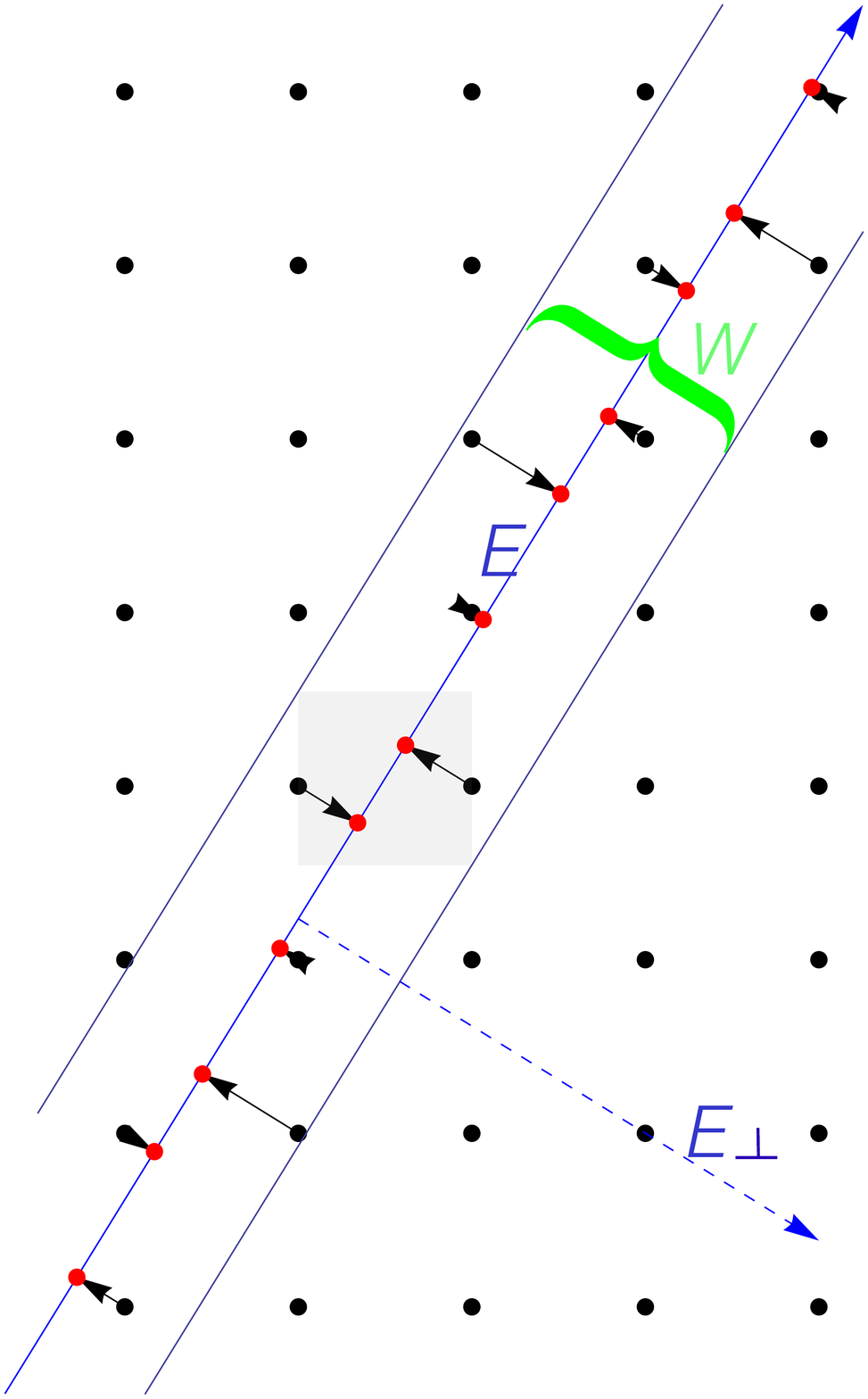}
 }

\caption{(Color online)  Two versions of the projection method: (a) Using Voronoi regions, shown as shaded (green) boxes.
The arrows indicate the lattice points in $L \subset \RR^2$ which are projected onto the line $E$ with irrational slope. (b) Canonical version,
projecting lattice points lying inside a strip.
  \label{fig:projectionMethod1} 
}
\end{figure}

In one version of the projection method \cite{93}, we choose which lattice points to project by considering
the Voronoi region of a lattice  point $p$ of the cubic lattice $L$, i.e., a cube centered at $p$,
and projecting onto $E$ exactly those lattice points $p$ whose Voronoi regions intersect $E$; 
see fig.~\ref{fig:projectionMethod1}(a).
This gives a set $\Lpar$ of points in the subspace $E$, which 
is a quasiperiodic lattice if $E$ is \emph{totally irrational}, i.e., the angle made with each lattice direction is an irrational multiple of $\pi$ \cite{93}.

The Voronoi projection method in fact produces the same quasilattice as the following canonical projection method \cite{93}: 
consider a cube whose side length is the lattice spacing, centred at the origin, and project it onto the orthogonal subspace $\Eperp$,  giving a set $W$.
The lattice points in $L$ which lie inside $W \times E$ are projected onto $E$; see fig.~\ref{fig:projectionMethod1}(b).

\paragraph{Construction of embedding:-}
The quasiperiodic LG consists of balls of dimension $m$ in $E$ centred at each point of $\Lpar$.
We will, in a sense, reverse the projection method, to construct a billiard obstacle $K$ inside  a unit cube $C$ of $\RR^{n}$ with periodic boundary conditions, designed specifically such that
trajectories of the billiard dynamics \emph{parallel} to $E$ inside $C$   automatically give trajectories of the quasiperiodic LG when the dynamics is ``unfolded'' 
to the whole of $\RR^{n}$ -- i.e., by following the dynamics using periodic boundary conditions in one cell, but keeping track of which cell has been reached \cite{CM06}.
%
%
%

We start by considering free motion of particles with arbitrary initial positions inside $C$, but whose velocities are constrained 
to move \emph{parallel} to $E$. When they cross the periodic
boundaries, they keep the same velocity, but jump to a different translated copy, or ``slice'', $E_{\vv} := E + \vv$, parallel to $E$ but translated by a vector $\vv$ which is orthogonal to $E$, i.e., such that $\vv$ is
in $\Eperp$. The periodic boundary conditions can thus be thought of as allowing us to ``wrap'' the whole of the plane $E$ in $\RR^{n}$ inside one single unit cell $C$.

Since $E$ is totally irrational,
a trajectory emanating from a generic 
initial condition in the cube $C$ will fill $C$ densely \cite{116}; 
similarly, the collection of parallel slices $E_\vv$ which are visited by a given trajectory also fill $C$ densely.
We will construct the billiard obstacle $K$ inside $C$ such that the dynamics  retains this denseness property.

Since $E$ is  wrapped inside $C$, 
we can  think of $C$ as corresponding to \emph{each} of the Voronoi cells in $\RR^{n}$ which intersect $E$.
The projection method then  requires that we project the origin onto each  slice $E_{\vv}$ which intersects $C$. We denote by $W$ the set of all such points; it turns out to be exactly the projected set used in the  canonical projection method. Note that 
 $W$ is a subset of full dimension inside $\Eperp$, which can be constructed in general by
projecting the vertices of the cube $C$ onto $\Eperp$ and taking the convex hull of the resulting points.

\begin{figure}
\includegraphics[scale=0.9]{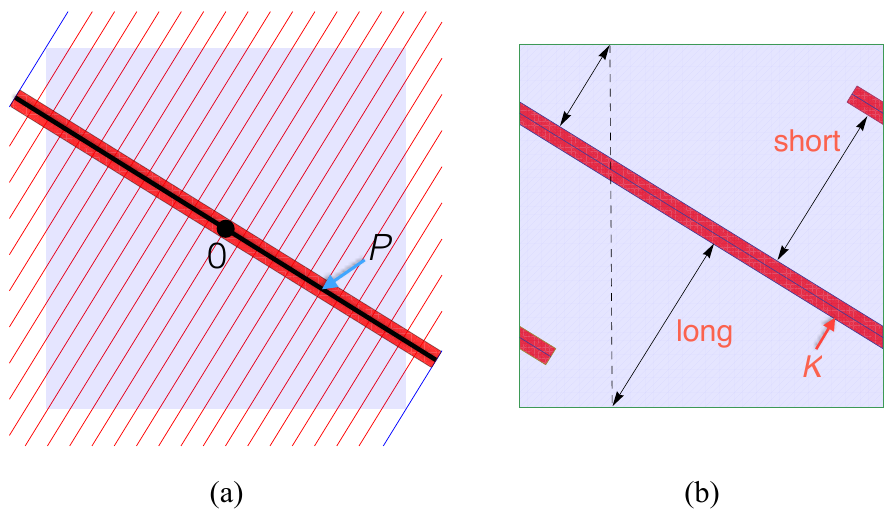}
 \caption{Embedding of a 1-dimensional quasiperiodic lattice into a two-dimensional periodic billiard. (a) Construction of the set $P$ by projection and ``thickening''. The parallel diagonal lines (red online) show different ``slices'' $E_{\vv}$.  (b) The billiard obstacle $K$, after periodizing $P$, shown as the oblique (red) bar. Examples of short and long paths for the resulting billiard dynamics are shown. 
 \label{fig:quasiperiodic1d}
}

%

\end{figure}

To complete the construction, 
we now define a set $P$ as the result of projecting not just a single point at the origin, but rather the whole billiard obstacle in $E$, an $m$-dimensional ball $B$ of radius $r$ centred at the origin, so that $P := W \times B$. However, some parts of $P$ constructed in this way  fall outside  $C$; see fig.~\ref{fig:quasiperiodic1d}(a).
We finish by periodizing $P$ via the periodic boundary conditions, giving a new set $K$, also of dimension $n$, which is the billiard obstacle that we finally use; see fig.~\ref{fig:quasiperiodic1d}(b).
%
%
%
Since the billiard particles move parallel to $E$, and the boundary of the obstacle $K$ is cylindrical, with axis perpendicular to $E$,
the particles will always move parallel to $E$ in $\RR^n$. When the resulting billiard trajectory is unfolded to $\RR^n$, it then gives rise
to a billiard trajectory  in the quasiperiodic LG.

As a simple example, consider a one-dimensional quasiperiodic billiard with $n = 2$ and $m = 1$, so that $C$ is a square, $E$ is a straight line through the origin with 
irrational slope $\alpha$, and the obstacle $B$ becomes a line segment of length $r$. 
The set $K$ then consists of three thickened line segments with slope $-1/\alpha$, as shown in 
figure~\ref{fig:quasiperiodic1d}(b).
 Since $K$ divides the square completely into two parts, there is, as expected, no diffusion in this case: any given trajectory remains confined, bouncing between two neighboring obstacles. However, this point of view easily allows us to calculate, for example, the relative probabilities of long and short paths starting from random initial conditions, by calculating the size of the respective areas.

\paragraph{2D quasiperiodic Lorentz gas:-}

\begin{figure}
 \centering
 \includegraphics[scale=0.3]{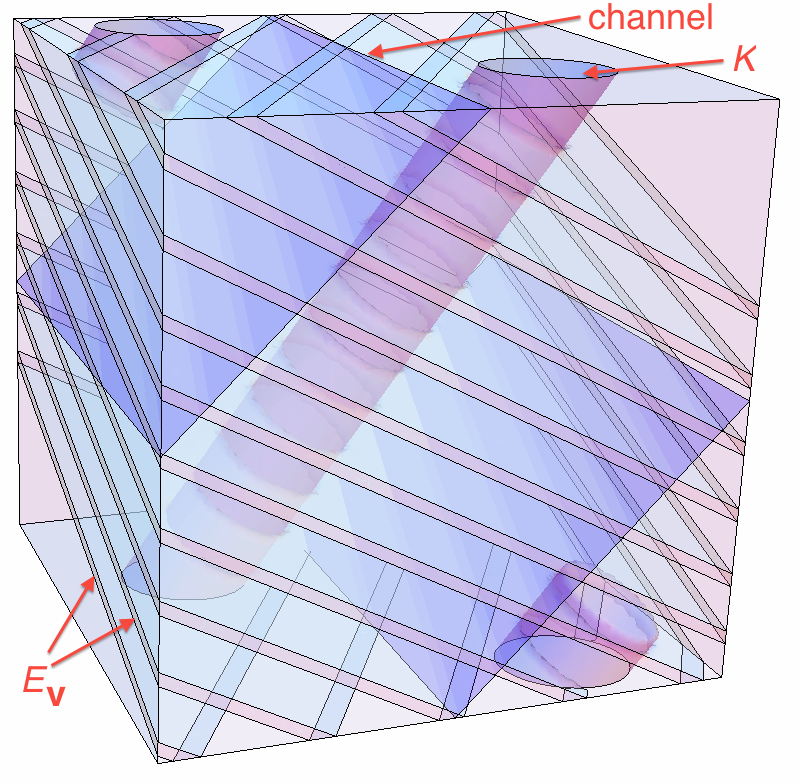}
\caption{QLG en 2D embedded in a 3D cube. The three pieces of cylinder form the billiard obstacle $K$, and the billiard dynamics takes place in the planes $E_{\vv}$ orthogonal to $K$.
There are two  planar channels (not shown explicitly) lying along the cube's faces, and an additional channel shown at an angle of $\frac{\pi}{4}$ which does not intersect $K$ for this choice of obstacle radius.
  \label{fig:Quasicrystal}
}
\end{figure}

We now turn to the simplest non-trivial application of this construction,  a 2D quasiperiodic LG formed by projecting
a 3D simple cubic lattice with cubic unit cell $C$.  The subspace $E$ is a totally irrational plane through 
the origin,  and the obstacle $B$ is a disc of radius $r$. The resulting 3D periodic billiard model is shown in figure
 \ref{fig:Quasicrystal}; the corresponding billiard obstacle $K$ consists of three segments of a cylinder, in analogy to the previous section. The two parts of the cylinder which come from periodizing $W$ are capped by those planes $E_{\vv}$ which pass through the respective vertices of $C$.

 The construction provides a natural way to interpret a uniform distribution of initial conditions in the phase space of the quasiperiodic LG: positions are uniform in the cube $C$, and velocities have unit speed and point in uniform directions parallel to $E$.
 We conjecture that the dynamics is ergodic, i.e.,  from almost any starting point outside the billiard obstacle $K$ (except for a set of measure zero), the corresponding trajectory fills densely and uniformly 
 this phase space. Averages are then taken with respect to this uniform distribution.
  
%
%

\paragraph{Channels:-}
Motivated by the results on periodic Lorentz gases, we ask if it is possible that there exist channels through the quasiperiodic LG structure, which could lead to super-diffusive behavior \cite{8, 100}. The absence of periodicity leads to the expectation that this is not possible, as suggested in ref.~\cite{16}.
However,  our construction in fact shows that   quasiperiodic LGs \emph{do} have channels  when the obstacle radius is sufficiently small.

Indeed, suppose that the radius $r$ of the obstacle $S$, and hence of the cylinder $K$, is small enough that none of the three cylindrical pieces 
(figure~\ref{fig:Quasicrystal}) intersect the remaining faces of the cube (i.e., those not intersected by their axes).
Then in the 3D billiard, there are ``planar channels'' $\Pi$ \cite{8} lying along each of these faces.
Restricting the dynamics to be parallel to the plane $E$, we see that 
each planar channel $\Pi$ in the 3D periodic billiard induces a rectangular channel in the 2D quasiperiodic LG, given by the intersection
of $E$ with $\Pi$.   Elsewhere we will show that this result generalizes to quasiperiodic LGs in $m > 2$ dimensions, which have $(m-1)$-dimensional ``principal horizons'' \cite{100}.
%
Note that as the radius decreases, additional channels may appear, as is also the case in the periodic LG \cite{89, 90}. For example,
fig.~\ref{fig:Quasicrystal} shows such additional planar channels oriented at an angle $\pi/4$.

\paragraph{Numerical results:-}

The above construction provides a direct and simple simulation method for quasiperiodic systems, 
by simulating directly billiard dynamics in the higher-dimensional system with  obstacle $K$ and periodic boundary conditions.
We applied this to the 2D quasiperiodic LG to investigate its diffusive properties as a function of the radius $r$ of the billiard obstacles. 

 In our simulations, we place $10^6$
particles distributed with uniform positions in the three-dimensional cube $C$, and velocities with unit speed distributed with  uniform directions parallel to the subspace $E$, and calculate the mean-squared displacement (MSD) $\langle x(t)^2 \rangle$ as a function of time.  The subspace $\Eperp$ was taken aligned along the unit vector $(\frac{1}{\phi+2}, \frac{\phi}{\phi +2}, \frac{\phi}{\sqrt{\phi + 2}})$, where $\phi := \frac{1 + \sqrt{5}}{2}$ is the golden ratio.

\begin{figure}[t]
  \centering
  \subfigure[]{
\includegraphics[scale=0.3, angle=-90]{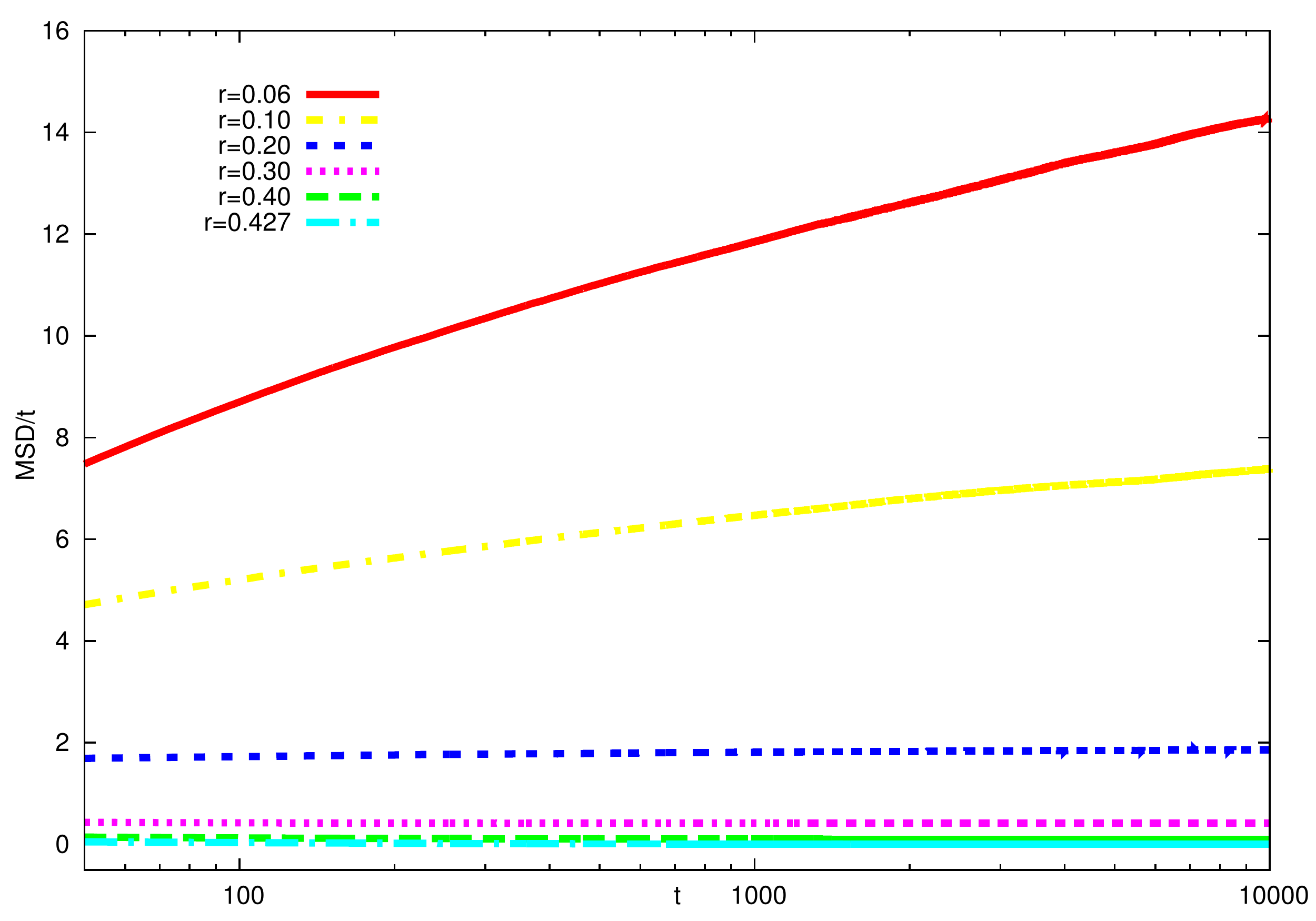}
  }
 \subfigure[]{
 \includegraphics[scale=0.3, angle=-90]{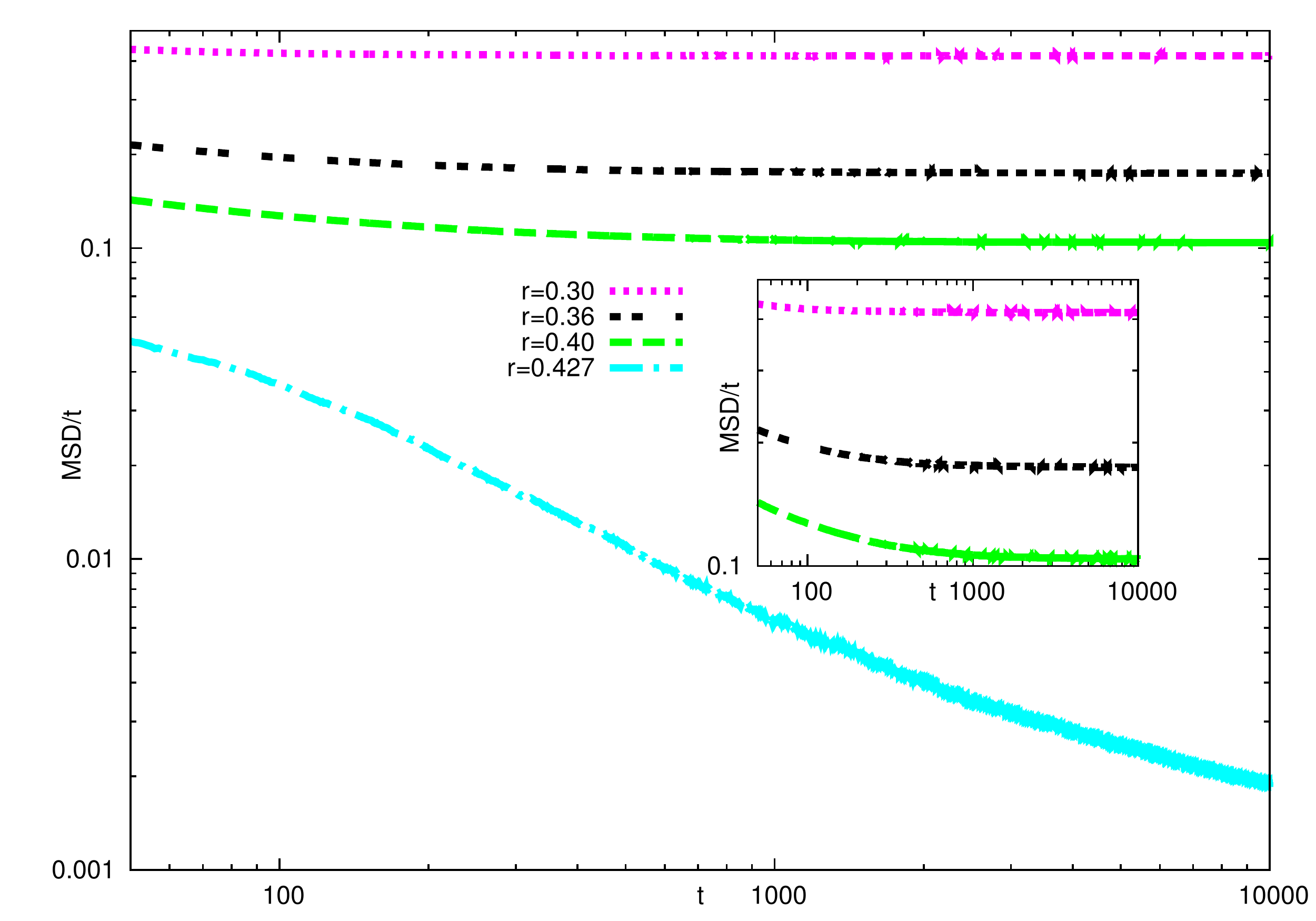}
 }

 \caption{${\langle x(t)^2 \rangle} / {t} $ as a function of $t$ for  different radii $r$, shown in  (a) semi-logarithmic and (b) logarithmic scale, with a zoom in the inset.
 }
   \label{fig:diffusion}
 \end{figure}

 Figure~\ref{fig:diffusion} shows $\langle x(t)^2 \rangle / t$, in order to emphasize deviations from normal diffusion.
We  find that for small radii, this quantity increases in time, so that super-diffusion is present. However, the correction does not appear to be a simple logarithmic factor, as for periodic LGs with channels, but rather seems to be fit well by expressions of the form $ \langle x(t)^2 \rangle \sim t (\ln t)^\delta$, with an exponent $\delta<1$ which depends on the radius $r$.

This super-diffusive behavior occurs due to the presence of channels.
%
As $r$ increases, the channels are blocked one by one, as the billiard obstacles expand sufficiently to intersect the planes defining the channels. Note that once the billiard obstacles are large enough to cross the faces of the cube, their periodic images must also be taken into account as additional obstacles in the simulations; for this reason, the algorithm is most efficient when the obstacles are small enough for this not to occur, which is exactly the most  difficult case to treat 
using standard methods, for which very large quasiperiodic lattices must be constructed.

At a certain critical radius around $r \simeq 0.21$ (discussed in more detail elsewhere), all of the planar channels become blocked, but the billiard obstacles still do not touch each other. This is the equivalent of the ``finite horizon'' regime in periodic LGs \cite{8, 100}, and the simulation results for the quasiperiodic LG indeed appear to show normal diffusion, with $\langle x(t)^{2} \sim t$, in this case. However, as $r$ increases further, there is another critical radius around $r \simeq 0.30$, at which the billiard obstacles touch and then begin to overlap one another. In this regime we find \emph{sub}-diffusion (highlighted in  fig.~\ref{fig:diffusion}(b)) as in the random, overlapping LG, until the final critical radius $r \simeq 0.43$ is reached, when all particles become localized in bounded regions of space and there is no longer diffusion.

%

\begin{figure}[t]
\centering
 \subfigure[]{
 \includegraphics*[scale=.21]{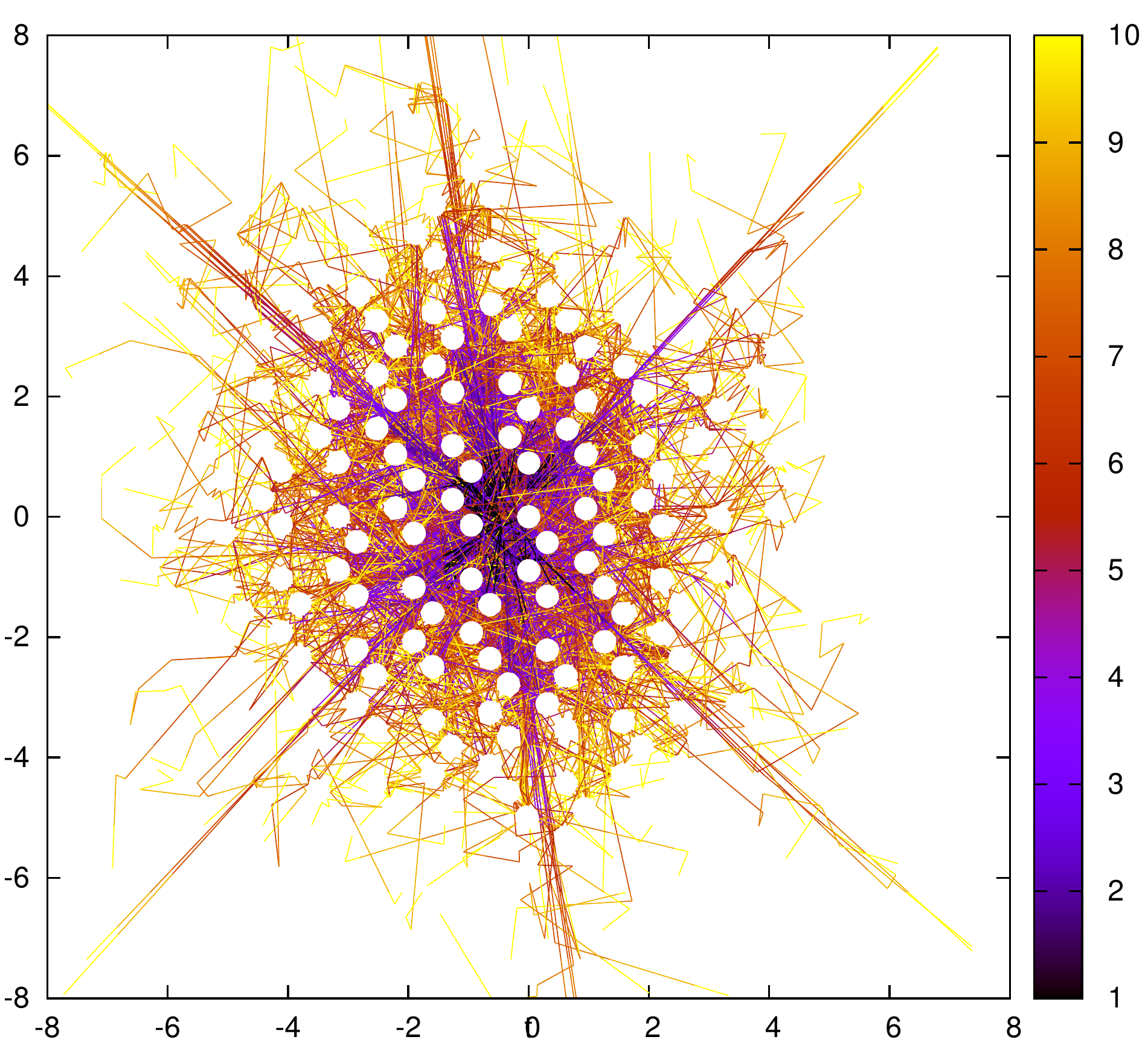}
 }
 \hspace*{-5pt}
 \subfigure[]{
\includegraphics*[scale=0.21]{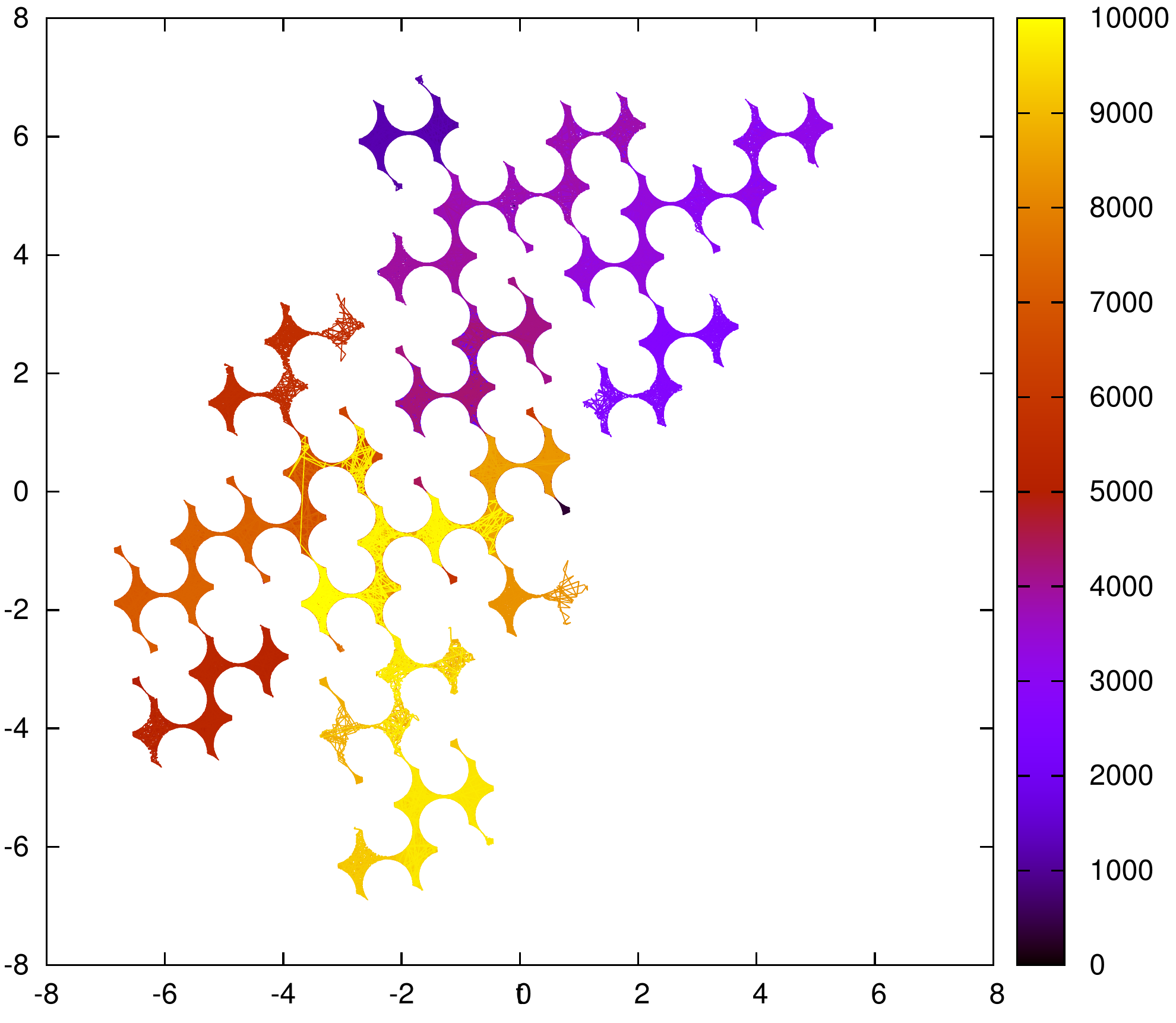}
  }
\caption{Representative trajectories in the 2D quasiperiodic LG; the gray scale (color online) indicates time, as shown in the respective color bars. Blank, circular regions correspond to the billiard obstacles; the positions of the obstacles are identical in both figures. (a) $r=0.2$;  several initial conditions are shown.  (b) $r=0.425$; single initial condition.
 \label{fig:Trajectories}
}
\end{figure}

Figure~\ref{fig:Trajectories}(a) shows  several representative trajectories for small $r$ in the super-diffusive regime,
 highlighting the channels  within the quasiperiodic LG which give rise to the super-diffusive behavior. For the radius $r$ shown, there are three directions in which channels occur, corresponding to those in fig.~\ref{fig:Quasicrystal} -- two along two perpendicular faces of the cube, and one at an angle of $\frac{\pi}{4}$.  
 Figure~\ref{fig:Trajectories}(b) shows a single trajectory for large $r$, close to the percolation threshold at which diffusion ceases.
Narrow bottlenecks between cavities of different sizes are visible, which explains the observed sub-diffusive behavior,
in analogy to the overlapping random Lorentz gas. 
 
In summary, we have introduced a construction to study quasiperiodic lattices by  embedding them into a periodic unit cell in a higher dimensional space.
This point of view transparently shows that there  quasiperiodic Lorentz gases have channels for  small obstacle radius, and provides a simple, direct and efficient simulation method for dynamics in quasiperiodic structures. Numerical simulations exhibit three diffusive regimes, including super- and sub-diffusion.
We expect that this construction 
can be profitably applied to simulate more complicated quasiperiodic systems.


The authors thank Domokos Sz\'asz for useful discussions at the Centro Internacional de Ciencias, in Cuernavaca, Mexico. Financial support is acknowledged from  CONACYT for ASK's doctoral studentship, and from the SEP-CONACYT grant CB-101246.


\end{document}